\documentclass[10pt, conference, compsocconf]{IEEEtran}
\usepackage{cite}
\usepackage{amsmath,amssymb,amsfonts}
\usepackage{algorithmic}
\usepackage{graphicx}
\usepackage{textcomp}
\usepackage{xcolor}
\usepackage{tikz}
\usepackage{pgfplots}
\usepackage{subcaption}
\usepackage{url}
\usepackage{listings}
\usepackage{mathtools}
\usepackage{multirow}

\usepackage[frozencache,cachedir=.]{minted}
\setminted[]{
    xleftmargin=.20cm,
    xrightmargin=.05cm,
    frame=single,
    framesep=.25cm,
    tabsize=1,
    breaklines,
    fontsize=\scriptsize
}

\newcommand{\inlinedsection}[2][6pt]{\vspace{#1}\noindent\textbf{#2}.}

\ifCLASSINFOpdf
\else
\fi
\begin{document}
\title{Secure IoT Data Analytics in Cloud via Intel SGX}


% \author{
% \IEEEauthorblockN{Authors Name/s per 1st Affiliation (Author)}
% \IEEEauthorblockA{line 1 (of Affiliation): dept. name of organization\\
% line 2: name of organization, acronyms acceptable\\
% line 3: City, Country\\
% line 4: Email: name@xyz.com}
% \and
% \IEEEauthorblockN{Authors Name/s per 2nd Affiliation (Author)}
% \IEEEauthorblockA{line 1 (of Affiliation): dept. name of organization\\
% line 2: name of organization, acronyms acceptable\\
% line 3: City, Country\\
% line 4: Email: name@xyz.com}
% }

%%% shihab: because of double-blind reviewing process, author names are hidden. Fix this in the camera-ready paper.

\author{
\IEEEauthorblockN{Md Shihabul Islam}
\IEEEauthorblockA{
\textit{Department of Computer Science}\\
\textit{The University of Texas at Dallas}\\
Richardson, TX 75080, U.S.A.\\
md.shihabul.islam@utdallas.edu
}\\
\IEEEauthorblockN{Latifur Khan}
\IEEEauthorblockA{
\textit{Department of Computer Science}\\
\textit{The University of Texas at Dallas}\\
Richardson, TX 75080, U.S.A.\\
lkhan@utdallas.edu
}
\and
\IEEEauthorblockN{Mustafa Safa Ozdayi}
\IEEEauthorblockA{
\textit{Department of Computer Science}\\
\textit{The University of Texas at Dallas}\\
Richardson, TX 75080, U.S.A.\\
mustafa.ozdayi@utdallas.edu
}\\
\IEEEauthorblockN{Murat Kantarcioglu}
\IEEEauthorblockA{
\textit{Department of Computer Science}\\
\textit{The University of Texas at Dallas}\\
Richardson, TX 75080, U.S.A.\\
muratk@utdallas.edu
}
}


%Telephone: (800) 555--1212, Fax: (888) 555--1212}


\maketitle

\begin{abstract}
The growing adoption of IoT devices in our daily life is engendering a data deluge, mostly private information that needs careful maintenance and secure storage system to ensure data integrity and protection. Also, the prodigious IoT ecosystem has provided users with opportunities to automate systems by interconnecting their devices and other services with rule-based programs. 
The cloud services that are used to store and process sensitive IoT data turn out to be vulnerable to outside threats. Hence, sensitive IoT data and rule-based programs need to be protected against cyberattacks.  
To address this important challenge, in this paper, we propose a framework to maintain confidentiality and integrity of IoT data and rule-based program execution. We design the framework to preserve data privacy utilizing Trusted Execution Environment (TEE) such as Intel SGX, and end-to-end data encryption mechanism. We evaluate the framework by executing rule-based programs in the SGX securely with both simulated and real IoT device data.
% evaluation summary %

\end{abstract}


% \fi
\IEEEpeerreviewmaketitle

\section{Introduction}
\label{sec:intro}

The Internet of Things (IoT) has transformed the way we live and work with their ubiquitousness, inexpensiveness, and convenience of usage. Increasingly, IoT devices are found in our day-to-day life, such as in smart homes, industrial automation, agriculture, smart transportation, healthcare etc. They have become a fundamental part of the modern society and still offer plenty of opportunities to make our life more comfortable and constructive. The recent growth of IoT is astonishing and it is predicted that there will be $64$ billion IoT devices by $2025$~\cite{ bussiness_insider_report}.
% ^ Example of iot device? %
Although, IoT  systems have many benefits, there are also plethora of security and privacy concerns related to IoT. IoT deals with vast amounts of highly vulnerable and sensitive data, which needs careful maintenance and secure storage and processing system to ensure user privacy and data protection. For instance, while listening for commands, Samsung’s Smart TV captures every words of its users no matter how private the conversation and transmits to a third party for conversion of speech to text~\cite{cnn_report_tv}. 

Recent development of cloud computing has provided the opportunity to use cloud-based services to collect, process, analyze, and mine large amounts of data~\cite{khan2000audio, wang2006automatic, petrushin2007multimedia, abrol2010twinner, wang2004automatic, awad2007web, awad2008predicting, nessa2008software, yen2002component}, which is both cost effective and less time consuming~\cite{shaon2017sgx}. A recent study found that, out of 81 common IoT consumer devices, $72$ send data to third parties, and rest to the original device manufacturer~\cite{avast_report, ren2019information}. Although, the cloud service providers ensure that data is always protected at rest, they are vulnerable to many security threats during transmission, and computation~\cite{masud2011data, masud2007hybrid, thuraisingham2008data, al2019bimorphing}; e.g., data breaches, especially in the public cloud services~\cite{iot_vul_report}. For instance, CloudPets, which manufactures smart stuffed toys for children, stored all the data (i.e., email, password, photos, voice recordings) in the unsafe cloud, exposing over $820,000$ user accounts including $2.2$ million voice recordings~\cite{cnn_report_toy}. In addition, adversaries may physically access the machines or obtain root privileges of the machines deployed at the service providers’ premises and thus steal sensitive information with ease~\cite{ayoadeevolving}. 

Moreover, the availability of cheap yet powerful IoT devices has paved the way for platforms to enable information passing among IoT devices and online services to automate different processes. These platforms, such as Samsung’s SmartThings~\footnote{\url{https://www.smartthings.com}} and  IFTTT (If-This-Then-That)~\footnote{\url{https://ifttt.com}}, offer users to automate their smart home or industrial system through customized policy-based rules that control the interactions between devices. For example, to conserve energy and reduce cost, a user may program a rule that automatically turns off the air conditioner and the light bulbs when the user is away from home. However, this enlarges potential attack surface and privacy risks, since these automation policies and sensitive device information are shared with untrusted parties over the internet.
For instance, suppose we have a temperature sensor which can open windows in a room. A temperature-related application can periodically check the room temperature and if the temperature is above a predefined threshold, then the sensor will open the window. Now, if an attacker can get access to the logic code in the cloud, he/she can change the value of the threshold, which could trigger the window opening action and cause a potential problem of break-in. Therefore, conventional security mechanisms of the cloud services need to be enhanced to thwart adversaries from stealing sensitive data and information.

%MK: Need to clarify the relationship of this paper and \cite{islam2019secure}

In this paper, we present a system that is established based on our previously proposed framework~\cite{islam2019secure}.
More specifically, in our previous work, we envisioned a system to securely store and process IoT information in a privacy-preserving manner by utilizing proper cryptography techniques and the Trusted Execution Environments (TEEs). In this work, we develop and empirically evaluate the envisioned framework to ensure the integrity and confidentiality of sensitive IoT data, private user information, and vulnerable automation policies in the untrusted cloud by performing rule-based analytics on a popular TEE called Intel Software Guard Extensions (SGX)~\cite{intel_sgx}.
Intel SGX creates an isolated secure memory container, where the code and data can be safely stored and executed. No adversaries, not even higher privileged software such as operating system (OS) or virtual machine manager (VMM) can access the contents of SGX. Therefore, our framework stores delicate IoT data, and user information in encrypted format, and securely executes rule-based interactions of IoT devices in the enclave, so that adversaries cannot manipulate or steal information.
Moreover, we ensure data security in transit from IoT devices to cloud service provider with SGX by following strong end-to-end encryption mechanism. That means, in transit data is always kept in encrypted form, except when it is in the SGX. We evaluate our framework for the IoT rule-based home automation setting with both simulated and real device data and study its efficacy in terms of both performance and security.

To summarize, in this paper, we propose the following contributions.
\begin{itemize}
  \item We propose and develop an end-to-end encrypted system for securely analyzing IoT data using TEEs, particularly Intel SGX.
  \item We perform thorough evaluations to assess the framework with both simulated and real IoT device data.
  \item We conduct security evaluations for potential vulnerabilities of the system.
\end{itemize}

The rest of the paper is organized as follows. Section~\ref{sec:background} presents some background on Intel SGX and IoT system. Section~\ref{sec:prob_statement} explains the problem statement and threat model. Section~\ref{sec:proposed_work} introduces our framework architecture and its components.
Section~\ref{sec:evaluation} describes the experiments and evaluation of the framework.
Section~\ref{sec:fwork} and Section~\ref{sec:rwork} describes future work and related work, respectively. 
Finally, Section~\ref{sec:conclusion} concludes our work.
\section{Background}
\label{sec:background}

\subsection{Intel SGX}
%Trusted Execution Environment (TEE) is a secure area inside the main processor designed to run in a parallel way with the operating system in an isolated environment. It ensures the software components loaded in the TEE are protected with respect to confidentiality and integrity and everything outside TEE is untrusted. Moreover, unpredictable software running in the main system software cannot affect software running in the TEE.
Intel's Software Guard Extensions (SGX)~\cite{intel_sgx} is one of the state-of-the-art Trusted Execution Environments (TEE), that provides hardware-assisted secure area of memory where trusted part of an application can be executed. This ensures the integrity and confidentiality of an application's security-sensitive computation and data on a computer where all the privileged software such as operating system is potentially malicious. With the help of SGX, application developers can protect their code and data from modification or disclosure by an adversary by creating a private memory region called Enclave and deploying those sensitive code and information within the Enclave.
The contents of enclaves are stored in the Enclave Page Cache (EPC), which is a piece of cryptographically protected memory with a page size of 4KB. Enclave is isolated from other processes or applications running at the same or higher privilege levels. No code, not even the higher privileged code such as Operating System (OS) or Virtual Machine Manager (VMM), can alter the contents of the Enclave, which makes it pretty robust from outside attacks and makes the attack surface
of the SGX as minuscule as possible~\cite{karande2017sgx}. 
%Figure~\ref{fig:attack_surface} demonstrates the dramatic difference between attack surfaces with and without the help of Intel SGX enclaves~\cite{intel2015tutorial}.

In SGX, a remote entity can cryptographically verify the integrity of an enclave and create a secure channel for sharing secrets with it. In Intel SGX architecture, this process is called \emph{Attestation}. Intel SGX guarantees protection of data when it is maintained within the boundary of the enclave. When the data needs to be stored outside the enclave, SGX encrypts the contents before writing to untrusted memory, so that integrity and confidentiality of data remains intact. The process of encrypting the data is called \emph{Sealing}. The data can be read back in by the enclave at a later date and then decrypted or unsealed. The encryption keys are derived internally on demand and are not exposed to the enclave.

% \begin{figure}[t]
% \centering
% \includegraphics[width=0.8\columnwidth]{figures/sgx_attack_surface.pdf}
% \caption{Attack-surface areas with and without Intel SGX enclaves.~\cite{intel2015tutorial}}
% \label{fig:attack_surface}
% \end{figure}

\subsection{IoT System and Security}
In an IoT system, a collection of smart devices and users communicate with each other to achieve a common goal in the industrial and commercial environments as well as in our personal life~\cite{sicari2015security}. IoT security refers to securing those connected devices and networks in the internet of things ecosystem. With cosmic IoT ecosystem, security threats are getting amplified and the IoT security must be designed to protect systems, networks, and data from a broad spectrum of attacks.
Specially, cloud-based services provide solutions to connect the IoT devices and collect data from the most sensitive and personal domains of our life to process, manage, and analyze the data utilizing different data mining and machine learning techniques~\cite{masud2009multi, haque2016efficient, masud2011detecting, masud2010classification, awad2004effective, al2012stream, breen2002image, masud2015systems}. These solutions must ensure data anonymity, confidentiality, and integrity as well as prevent unauthorized access to the system. 

%There already exist some solutions for the IoT. Amazon AWS IoT~\footnote{\url{https://aws.amazon.com/iot/}} offers a full solution to connect IoT devices, and collect, store, and analyze device data in the cloud. IBM Watson IoT Platform~\footnote{\url{https://www.ibm.com/us-en/marketplace/internet-of-things-cloud}} is a cloud-hosted service that connects, manages and secures IoT devices with data management and analytics services. Microsoft Azure IoT~\footnote{\url{https://azure.microsoft.com/en-us/overview/iot/}} is another powerful platform to build industrial solutions and process massive quantities of data from all kinds of IoT devices using AI and machine learning in the cloud. Mozilla WebThings~\footnote{\url{https://iot.mozilla.org/}} is an open platform for monitoring and controlling devices over the web.

There already exist some solutions for the IoT, such as Amazon AWS IoT~\footnote{\url{https://aws.amazon.com/iot/}}, IBM Watson IoT Platform~\footnote{\url{https://www.ibm.com/us-en/marketplace/internet-of-things-cloud}}, Microsoft Azure IoT~\footnote{\url{https://azure.microsoft.com/en-us/overview/iot/}}, Mozilla WebThings~\footnote{\url{https://iot.mozilla.org/}} and so forth.
Even though, these solutions offer some level of security related to data~\cite{masud2008cloud, parveen2011insider, lavee2007framework, tu2008secure, abedin2006detection}, they are highly dependent on users' trust towards their platform. The users trust these services with their private data and an unfortunate event of compromised cloud could endanger the privacy and confidentiality of user data~\cite{daubert2015view}. Therefore, we need a more robust strategy and technique to protect the data in both trusted and untrusted cloud environments.

%MK: Any SGX, IOT work, we need to discuss them.
\subsection{Automation using IoT}
One of the most powerful features of the IoT system is the ability to automate processes with the help of devices without any human intervention. The most obvious conveniences of the IoT automation are more operations, more accuracy, and low cost.
Usually, IoT devices consist of embedded sensors and actuators, which help the devices to interact with the physical environment. Sensors can collect physical states, which are known as \emph{Events}. These events, such as temperature reading, dust level, or door lock state, are sent to the cloud or hub for further processing. Afterwards, based on user-defined protocols and event data, appropriate action commands are sent to the device actuators. Generally, to transfer data between devices and cloud/hub, suitable protocol is used that supports limitations of the environment such as low powered devices. There are some IoT programming platforms such as Samsung’s SmartThings, IFTTT, Apple’s HomeKit~\footnote{\url{https://www.apple.com/ios/home/}}, Zapier~\footnote{\url{https://zapier.com}}, openHAB~\footnote{\url{https://www.openhab.org}} etc. that provide app-specific services of controlling and managing devices, data collection, and  device interactions. They also provide tools that allow developers to write applications and automations through various APIs~\cite{celik2019iotguard}.

One of the most widely approved IoT programming platforms, especially for the home automation, is the rule-based \emph{Trigger-Action} platform. This platform allows users to create custom simple and complex automations on services through rules that operate on the cloud. More specifically, the trigger-action rule platform performs some actions when a certain trigger event takes place. Typically, users define the rule by connecting a trigger-event in a service and an action-command in a separate service. When a device event matches the trigger-event, the appropriate action-command will be fired on the relevant service. 
For example, a user may define a rule: \emph{Turn on the hall lights if motion is detected on the lawn}. Here, the trigger event is the detection of motion by the motion sensor and the action command is to turn on the lights using a smart switch. Triggers may contain trigger properties that determine under what circumstances the trigger event should occur. Similarly, action commands have action properties which are the parameters of the action~\cite{bastys2018if}. These rule-based platforms are substantially benefiting smart home and industry automation systems. For instance, IFTTT has a community of $11$ million users running over $1$ billion rules each month with over $600$ partner services~\cite{ifftt_report}.

% Different rule types?
\section{Problem Statement \& Threat Model}
\label{sec:prob_statement}

\subsection{Problem Statement}
The use of rule-based platforms to control and interact with IoT devices is a powerful tool, but without proper and thorough security measures, it could lead to various unsafe conditions and unrecoverable loses. Generally, IoT devices expose three categories of information: \emph{Stored Data} (i.e., device identifiers, user identifiers, activity logs), \emph{Sensor Data} (i.e., information or physical states obtained from the environment by the sensors of devices), and \emph{Activity Data} (i.e., information about how the devices are used via automation rules or user interaction)~\cite{ren2019information}.
These information may be shared with two kinds of party: \emph{First party} and \emph{Third party}. First party includes the manufacturer of the IoT devices that are responsible for the device functionalities. On the other hand, Third parties are the organizations providing computing resources such as cloud providers or analytics companies. IoT devices expose those three types of data explicitly with these parties, which could pose potential data privacy issues. 

In the IoT ecosystem, as the service providers are trusted with abundant user information, a major challenge arises in the form of balancing trust in these service providers and need for privacy. Although, the cloud service providers ensure that data is always protected at rest, during transmission, and computation; in reality they are vulnerable to many security threats, e.g., data breaches, especially the public cloud services~\cite{iot_vul_report}. In addition, severe lack of proper encryption techniques could expose sensitive information about the users. As a consequence, significant privacy risks could emerge as malicious third party services can track information about users for monetary purposes as well as learning user activities within homes.
For instance, smart speakers in home can covertly record user conversations without permission and stream it to other users or parties~\cite{alexa_report}. 
Moreover, adversaries may physically access the machines deployed at the service providers premises or obtain root privileges of the machines  by taking advantage of weak access control mechanism and thus steal sensitive information with ease.
%For instance, suppose we have a temperature sensor which can open windows in a room. A temperature-related application can periodically check the room temperature and if the temperature is above a predefined threshold, then the sensor will open the window. Now, if an attacker can get access to the logic code in the cloud, he/she can change the value of the threshold, which could trigger the window opening action and cause a potential problem of break-in. Therefore, only security policy-based rules or operating system access control mechanisms of the cloud services are not sufficient to thwart adversaries from stealing sensitive data and information.

\subsection{Threat Model}
In this paper, we consider an adversary that seeks to surreptitiously gain insight into sensitive user information in the IoT system. More specifically, the adversary tries to access IoT device information and data stored in the cloud. The adversary tries to deploy a rule-level attack either by compromising the existing stored rule, by injecting malicious rule into the system, or by simply observing the rules to gain insight. Moreover, adversaries may eavesdrop the network traffic to retrieve information.
The principal objective of the adversary would be to obtain private information of the user, specially his/her surrounding environment such as in a smart home system. We assume that adversaries cannot get root access to the devices or compromise communication protocols. Denial-of-Service (DoS) attacks~\cite{iot_ibm_report} and protocol flaw attacks~\cite{ronen2017iot} are out of our scope.

\section{Proposed System Architecture}
\label{sec:proposed_work}

\begin{figure}[t]
\centering
\includegraphics[width=0.99\linewidth]{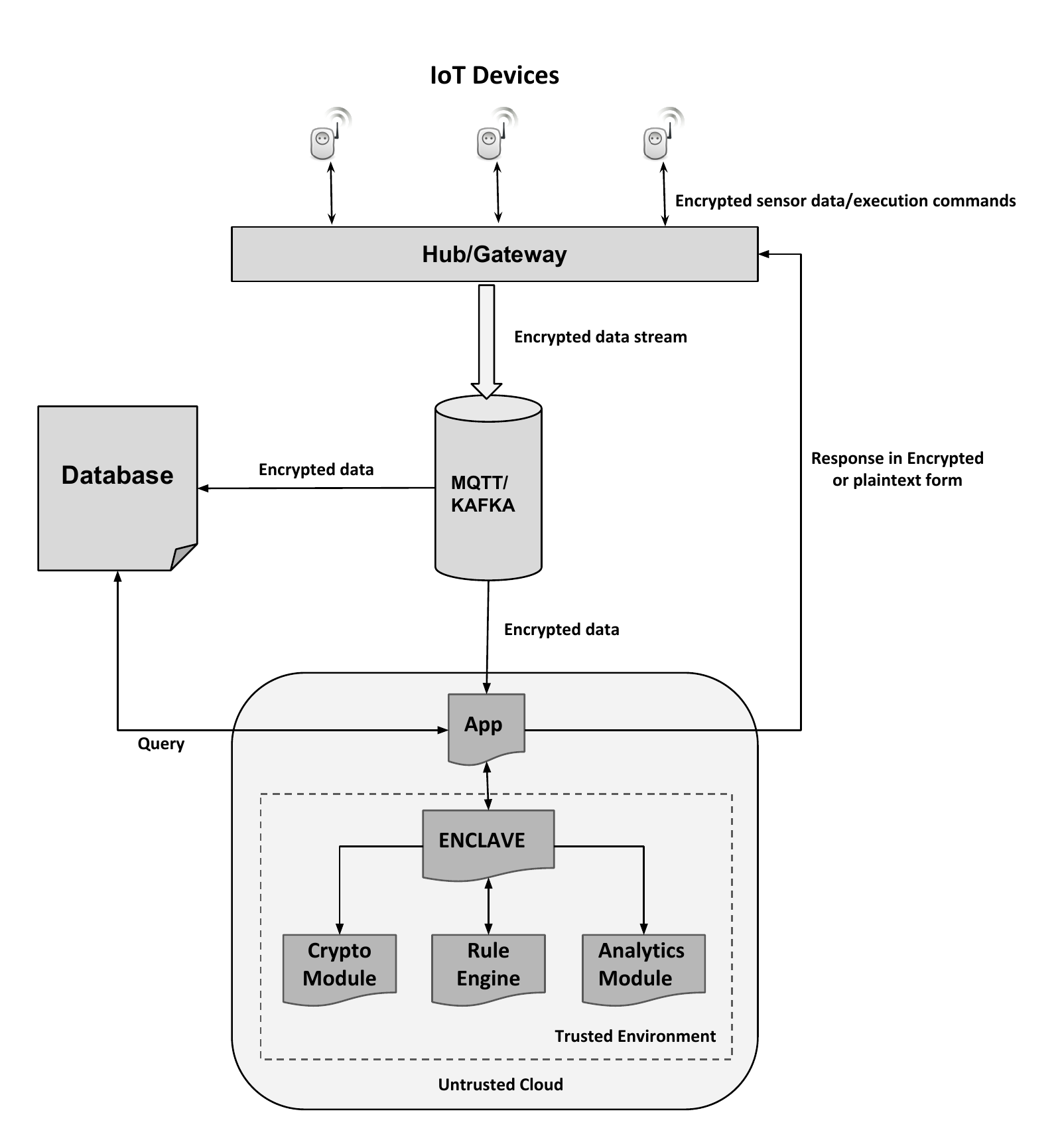}
\caption{Framework architecture.}
\label{fig:framework}
\end{figure}

Our aim is to develop a secure cloud-based end-to-end encrypted data analytics platform, especially designed for IoT setting. Our goal is to alleviate data security and privacy issues by utilizing proper cryptographic techniques and trusted execution environments such as Intel SGX. In this paper, we particularly focus on developing a rule-based secure IoT platform for smart home automation in the untrusted cloud. 

As discussed in section \ref{sec:background}, rule-based trigger-action platform is one of the most widely used IoT programming platforms in the world of IoT automation. Users provide trigger-action rules to automate their smart homes or smart industries leveraging the connectivity and ubiquitousness of IoT devices. These rules are stored and processed in the untrusted cloud platform or company-owned data silos, which poses a threat to the security and the privacy of the users. Moreover, lack of proper encryption techniques when communicating with untrusted cloud could expose sensitive information such as the identity of a device, user interactions with the device or private user information to eavesdroppers.
Therefore, in our framework, we aim to use Intel SGX to guarantee confidentiality and integrity
of sensitive data coming from IoT devices to untrusted remote platforms. 
%Intel’s Software Guard Extensions (SGX) is Intel’s latest iteration of a trusted hardware solution to the secure remote computation problem. The SGX design is centered around the ability to create an isolated container (Enclave), whose content receives special hardware protections that are intended to translate into confidentiality, integrity, and freshness guarantees. 
By utilizing SGX’s enclave features, we securely perform rule-based programming on delicate IoT data, so that no unauthorized personnel can unlawfully access data, user provided rules or any analytical results.

Usually, the required SGX enclave instances will be initialized by the cloud provider in the untrusted cloud platform. Once the enclave is initialized, it is expected to participate in a software attestation process, where it authenticates itself to a remote application server. Upon successful authentication, the application server is expected to disclose some secrets, in this case encryption/decryption keys, to the enclave on the untrusted platform over a secure communication channel.

The enclave in the cloud will communicate with the IoT devices in user homes via IoT gateways or hubs over HTTPS connection~\cite{ fernandes2016security}.  The communication protocol of HTTPS is encrypted with Transport Layer Security (TLS)~\cite{tls_report}, or formerly known as Secure Sockets Layer (SSL). In addition, to ensure end-to-end secure system, we use symmetric key encryption to communicate between the enclave in the cloud and the IoT hub. We use one of the most popular and widely adopted symmetric key encryption algorithms Advanced Encryption Standard (AES)~\cite{daemen1999aes} in our framework for this purpose. Hence, data in transit is always secure and eavesdropping on it is almost hopeless.

%%% Rule registration in sgx
To create an automation, a user first needs to register a trigger-action rule in the cloud via any web or app interface. For instance, in Samsung SmartThings, automation is created via SmartApps, which is essentially an AWS Lambda function or a WebHook endpoint ~\cite{ smartthings_smartapps}. SmartThings follows REST API architecture to control and communicate with SmartThings devices from the cloud~\cite{ smartthings_automation }.
We follow a similar architecture in our platform so that our framework is aligned with the well-established SmartThings system. We also adopt SmartThings JSON rule structure~\cite{ smartthings_rules}.

After registering the smart devices, users can define their rules for the automation of their devices. The rule contains a list of conditions for trigger and a list of actions for the desired operation. On one hand, the conditions specify the device events received from smart devices that triggers the rule. The device event could be a state of the device (i.e., switch on/off, door open/closed etc.) or a sensor reading of the device (i.e., temperature $90$F, dust level $20$ PM10, energy $130$ kwh etc.). On the other hand, the actions specify  what rules actually do. They are the commands sent to specific devices to control or actuate them in response of the defined trigger condition.  
Listing~\ref{listing_rule} presents a sample rule in JSON format. 
These rules are then sent to the untrusted cloud enclave after encrypting it. In our framework, rules will be safely stored in the database in encrypted form at all times and are only decrypted inside the SGX enclave, thus preventing the attacker from accessing or manipulating the rules. 

\begin{listing}[ht]
\begin{minted}{json}
{
	"name": "If user is home, set the thermostate mode to cool and turn on the lights",
	"ruleID": "hpGHOiCPPeE9-Nz8CYTO1Cmj5",
	"userID": "AI4gwcJ6I6DE-s5QwIgXXtm3q-oUbYddxRmCqe",
	"actions": [
		{
			"if":{
				"equals": {
					"left": {
						"device": {
							"devices": ["420CC6DD-5932-9DF4-945D4539"],
							"component": "main",
							"capability": "PresenceSensor",
							"attribute": "presence"
						}
					},
					"right": {
						"string": "present"
					}
				},
				"then": [
					{
						"command": {
							"devices": ["61902075-855B-4EF6-FE7AD97B"],
							"commands": [
								{
									"component": "main",
									"capability": "ThermostatMode",
									"command": "cool",
									"arguments": []
								}
							]
						}
					},
					{
						"command": {
							"devices": ["39A56C99-9A3A-45D7-D6537244"],
							"commands": [
								{
									"component": "main",
									"capability": "Switch",
									"command": "on",
									"arguments": []
								}
							]
						}
					}
				],
				"else": []
			}
		}
	]
}
\end{minted}
\caption{Sample Rule in JSON.}
\label{listing_rule}
\end{listing}

\begin{listing}[ht]
\begin{minted}{json}
{
    "deviceID": "420CC6DD-5932-9DF4-945D4539",
    "deviceEvents": [
        {
            "component": "main",
            "capability": "PresenceSensor",
            "attribute": "presence",
            "value": {
                "string": "present"
            },
            "unit": "",
            "data": []
        }
    ]
}
\end{minted}
\caption{Sample Device Event in JSON.}
\label{listing_device}
\end{listing}

%%% Data gather and rule-based decision making
The framework architecture is illustrated in Figure~\ref{fig:framework}. The IoT devices send device states or sensor values to the cloud via the hubs or gateways. The data is encrypted in the hub/gateway before sending to cloud and upon receiving a stream of such data from devices, SGX loads and decrypts the associated rules with the device in the enclave. As the system needs to deal with multiple data streams from various devices~\cite{de2016iot}, we use MQTT (Message Queuing Telemetry Transport)~\cite{mqtt}, which is designed as a lightweight publish/subscribe messaging transport, as our connectivity protocol. Additionally, note that, data is decrypted only inside the enclave using the secret key, which ensures data protection in transmission. Now, device event is compared with the condition of the rule (i.e., trigger) and generate corresponding response using action-command in the rule. This action-command is then encrypted and sent to the appropriate hub/gateway to control or actuate for the automation. The hub/gateway eventually takes care of transmitting the decision to the particular IoT device after decryption. Furthermore, users can define rules such that when the rule is triggered, user receives a notification instead of device actuation. 

For instance, Listing~\ref{listing_device} represents a sample device event received in the enclave. The event is generated from a \emph{Presence Sensor}. It contains the sensor attribute \emph{Presence} and current reading value, which is \emph{present}. After receiving the device event, the rule-engine in the enclave fetches from the cache corresponding rules for that device, in this case, the rule in Listing~\ref{listing_rule}. The rule-engine then proceeds to inspect the \emph{equals} condition of the rule. Here, the device attribute value and the rule condition value are the same, that is \emph{present}. Therefore, the rule is satisfied and will trigger the action commands, which are in the \emph{then} clause of the rule. These action commands will be sent to respective devices and executed there. In this example, a command will be sent to the \emph{thermostat} to set its state to \emph{cool} and another command to a smart \emph{switch} to set its state to \emph{on}.

To summarize, our framework ensures the integrity and confidentiality of IoT data and user rules and perform secure analytics by leveraging isolated memory containers such as SGX enclave. Moreover, we ensure data security in transit from IoT devices to cloud service provider with SGX by following robust end-to-end encryption mechanism of the data. That means, in transit data is always kept in encrypted form, except when it is in the SGX. Even if adversaries manage to steal the data in transit, they cannot reveal any information from it as it will be always encrypted.

\section{Implementation \& Evaluation}
\label{sec:evaluation}

%As the work is still in progress, in this section we present some preliminary experimental results. 
We evaluate the proposed framework by measuring computational time overhead of the whole process with simulated IoT data as well as data from real devices. In addition, we analyze memory access traces of the program to empirically evaluate the possibility of security threats due to an adversary that may analyze access patterns to encrypted data \cite{IKK12}. 

\begin{table}[]
    \centering
    \begin{tabular}{|c|p{0.13\linewidth}|p{0.1\linewidth}|p{0.09\linewidth}|p{0.1\linewidth}|p{0.065\linewidth}|}
    \hline
    \textbf{Case} & \textbf{Type} & \textbf{Ruleset size} & \textbf{Devices count} & \textbf{Total Device Events} & \textbf{Cache size} \\
    \hline
    \multirow{2}{*}{No SGX} & \multirow{5}{*}{Simulation} & 100 & \multirow{6}{*}{32} & \multirow{5}{*}{10000} & \multirow{6}{*}{100} \\ \cline{3-3}
    %\hline
     & & 400 & & & \\ \cline{3-3}
     %\hline
     \multirow{2}{*}{SGX} & & 1000 & & & \\ \cline{3-3}
     %\hline
     (w/o encryption) & & 5000 & & & \\ \cline{3-3}
     %\hline
     \multirow{2}{*}{SGX} & & 10000 & & & \\ \cline{2-3} \cline{5-5}
    %\hline
      & Real & 10 & & 1000 & \\
    \hline
    \end{tabular}
    \caption{Experimental setting}
    \label{tab:data_table}
\end{table}

\pgfplotstableread[row sep=\\,col sep=&]{
    interval & nosgx & sgxnoenc & sgx \\
    100     & 87.149 & 374.4  & 470.74\\
    400     & 112.44  & 510.05 & 565.54 \\
    1000    & 272.62  & 1310.69 & 1387.48 \\
    5000   & 1687.36  & 3661.89 & 3817.18 \\
    10000    & 3576.35 & 4650.31 & 4894.146 \\
    }\mydata

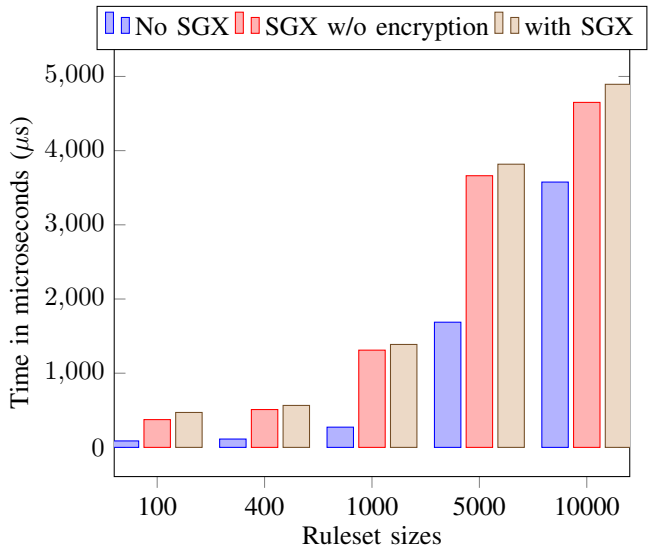
\begin{figure}
    \centering
    \begin{tikzpicture}
        \begin{axis}[
                ybar,
                symbolic x coords={100,400,1000,5000,10000},
                xtick=data,
                legend style={at={(0.5,1.1)},
	            anchor=north,legend columns=-1},
                ylabel={Time in microseconds ($\mu$s)},
                xlabel={Ruleset sizes}
                %nodes near coords align={vertical},
                %nodes near coords,
            ]
            \addplot table[x=interval,y=nosgx]{\mydata};
            \addplot table[x=interval,y=sgxnoenc]{\mydata};
            \addplot table[x=interval,y=sgx]{\mydata};
            \legend{No SGX, SGX w/o encryption, with SGX}
        \end{axis}
    \end{tikzpicture}
    \caption{Average execution time of the experiment based on $10000$ device events and ruleset sizes of $100$, $400$, $1000$, $5000$, and $10000$; performed for three cases: no SGX, SGX without encryption, and with SGX.}
    \label{fig:my_fig}
\end{figure}

\inlinedsection{Computational Evaluation}
Our goal of this evaluation is to measure the computational time overhead of the framework. More specifically, we want to discover how the integration of Intel SGX and the cryptographic techniques alter the time overhead of the process. For this experiment, we use simulated IoT data and rules in accordance with the Samsung’s SmartThings format (discussed in section \ref{sec:proposed_work}). The experimental setting is represented in Table~\ref{tab:data_table}. We consider three cases for the experiment: \emph{No SGX}, that provides no security guarantee of data; \emph{with SGX but without encryption mechanism}, which may provide integrity of data but lacks confidentiality; and \emph{with SGX}, that provides total security guarantee.
For SGX cloud, we use a system containing 8-core i7-6700 (Skylake) processor operating at 3.4GHz, running Ubuntu 18.04 with 64GB RAM, and client programs written in python (3.6) for simulating IoT devices.

At first, the client generates a set of rules $R=\{r_1, r_2, \dots, r_n\}$ for a set of his devices $D=\{d_1, d_2, \dots, d_m\}$. A device may have multiple rules associated with itself, e.g., $r_1$ and $r_2$ might both belong to $d_1$. Then, client encrypts the rules using AES encryption scheme, and sends to the SGX cloud. The SGX cloud loads the encrypted rules into the enclave and decrypts them. After parsing the rules, SGX enclave re-encrypts each rule separately with enclave's own secure key $k_{SGX}$ and stores it in a database as a $\langle$key, value$\rangle$ pair, where \emph{key} is the device ID and \emph{value} is the encrypted rules associated with that device. There's also a caching system (i.e., LRU, LFU) in the enclave to cache most frequently/recently used rules. Now, simulated IoT devices periodically send device states or sensor values to the cloud in encrypted form. Upon receiving this stream of encrypted data, the SGX loads the device data into the enclave, decrypts it, fetches associated rules from the database into the enclave, and decrypts the rules using $k_{SGX}$. It then generates the corresponding response using the triggers and actions specified in rules and sends the response back to the device after encrypting it. Upon receiving the response from SGX, device first verifies the integrity and the authenticity of the message. If both checks pass, device executes the message.
%MK: Can we change the orders in terms of security? Ie no SGX has no guarantee. SGX w/o encryption may provide integrity only. SGX provides both. 

Figure~\ref{fig:my_fig} represents the average execution time comparison for the three cases mentioned in Table~\ref{tab:data_table} for varying number of rules and $10000$ simulated device events. Needless to say, execution time of the experiment with SGX takes longer than the operation when we do not include SGX. Fortunately, the time execution overhead is not that significant.

Moreover, we perform a basic experiment with real IoT devices to evaluate the soundness of the system. We use sensor data from \emph{Foobot}~\cite{foobot} to control \emph{Philip Hue Bulb}~\cite{philips_hue} with some predefined rules in SGX. \emph{Foobot} is an indoor air quality monitor sensor, which can measure temperature, humidity, carbon dioxide level, volatile compounds in the air, and so on. Philip Hue Bulb is a smart bulb, which can be controlled with apps to turn on or off. At first, we store some predefined rules in the SGX enclave, where the trigger component of the rules involve \emph{Foobot} sensor values (i.e, temperature, humidity, and carbon dioxide level) and action component involve  changing the status of the \emph{Philip Hue} light bulb. Then, we periodically gather temperature, humidity and carbon dioxide level values from \emph{Foobot} and send to SGX cloud after encryption. Just like the above experiment, SGX enclave generates a response command according to the rule, which is then encrypted and sent to a python written program simulating the behavior of a hub. The response command is then decrypted and sent via https connection to the smart bulb to change the state. 

We observe the overall average execution and network delay time for the previously mentioned three cases with $10$ predefined rules and $1000$ device events (sensor values). We notice a similar result as before; \emph{SGX} incurring a slight time overhead with average overall time of $1.1\times10^5$ $\mu$s, where \emph{no SGX} and \emph{SGX without encryption} achieved $9.3\times10^4$ $\mu$s and $1.0\times10^5$ $\mu$s, respectively.

\inlinedsection{Security Evaluation}
As an adversary may obtain memory access traces of the program execution, s/he can infer sensitive information by analyzing access patterns from these traces, if the program displays distinguishing characteristics \cite{IKK12}. Therefore, our goal of this evaluation is to discover if the memory access traces of the program are indistinguishable or not. 
For this purpose, we use a randomly selected ruleset of size 10 and 3 set of device events (i.e., $S_1$, $S_2$, $S_3$) with each set containing 10 instances. Among these 3 sets, $S_1$ and $S_2$ are almost identical. We use Intel Pin tool~\cite{luk2005pin} to capture memory access traces (i.e., sequence of read and write operations) of the program executing in SGX simulation mode. We create probability distributions from these traces and use \emph{Kullback–Leibler divergence} (KL divergence) to differentiate between each traces. Table~\ref{tab:table} represents the comparison among the traces in terms of KL divergence score. From the table, we can observe that KL divergence score of near identical set $S_1$ and $S_2$ are lower than other two non-identical set comparison. As low KL divergence score means two distributions are more similar, we can deduce that almost indistinguishable data events create same memory access patterns in the experiment. As a result, this could be vulnerable to side channel attacks as adversaries can resend similar data continuously to the SGX cloud and observe memory access sequences to infer secrets from the enclave.

\begin{table}
\centering
\begin{tabular}{||c c||} 
 \hline
 Set Comparison & KL divergence score  \\ [0.5ex] 
 \hline\hline
 S\textsubscript{1} vs S\textsubscript{2}  & 0.199  \\ 
 \hline
 S\textsubscript{2} vs S\textsubscript{3} & 0.46  \\
 \hline
 S\textsubscript{1} vs S\textsubscript{3} & 0.39  \\ [1ex] 
 \hline
\end{tabular}
\caption{\label{tab:table} KL divergence score of memory trace distributions.}
\end{table}

\section{Limitations \& Future Work}
\label{sec:fwork}

Although Intel SGX is secure in design, it still suffers from pattern leakage attacks such as side channel attack as in ~\cite{intel_sgx}. These attacks, both memory level (shown in Section~\ref{sec:evaluation}) and network level, leak information and endanger data security. We plan to thwart such attacks by hiding the memory access patterns by introducing inconsistency in the side channel information~\cite{chandra2017securing} with injection of dummy data or by incorporating oblivious random access memory technique. Also, to make the security more robust, we plan to incorporate efficient access control mechanism that features decentralized authorization, protected permissions, and transitive permission delegation.
Furthermore, Intel SGX only supports a limited memory space (up to 128MB EPC) for data and code inside the enclave. This memory limitation calls for a distributed SGX system that will handle streaming data from IoT devices without any memory issues or sluggishness of the system. In the future, we aim to make our SGX system distributed, so that the framework do not face any unwanted memory issue or system slowdown. 
\section{Related Work}
\label{sec:rwork}

There has been some significant research on secure IoT data management over the past couple of years. \emph{Talos} stores IoT data securely in the cloud using cryptographic techniques and allows query processing over encrypted data~\cite{shafagh2015talos}. Even though the system is proved to be secure, the proof mainly depends on the robustness of the encryption algorithm as well as the application logic. In \cite{ayoade2018decentralized}, authors present a secure IoT data management system that uses a blockchain~\cite{nakamoto2008bitcoin}. They develop a decentralised framework that uses Ethereum smart contracts ~\cite{buterin2013ethereum} to control access permission of data, store audit trail of data access in the blockchain, and store raw data in encrypted form using Intel SGX. Although the framework is integrated with Intel Sgx and blockchain to ensure the security of the data, it does not handle any processing of the data securely. In addition, \cite{tahir2018using} utilizes Intel SGX to create enclaves that run virtual clones of physical IoT devices in the cloud to store, process, and share device generated data. 
\section{Conclusion}
\label{sec:conclusion}
As the usage of IoT devices increase, it is imperative that we protect sensitive user information and automation policy rules from malicious attacks. This paper proposes a framework that provides secure data analytics system by leveraging Intel SGX and strong cryptographic techniques. We execute basic trigger-action rule-based program for automation in the SGX enclave to ensure user privacy, data integrity and confidentiality. Moreover, strong encryption mechanism guarantees data privacy in transit and storage, making the system end-to-end encrypted. We evaluate the proposed framework by using data from simulated and real IoT devices, by performing rule-based decision making inside SGX enclave securely, and show that the overhead due to encryption and SGX based processing is not significant.



\section*{Acknowledgment}
%The research reported herein was supported in part by NIH award 1R01HG006844, NSF awards CICI-1547324, IIS-1633331, CNS-1837627, OAC-1828467, and ARO award W911NF-17-1-0356.

%NSF awards DMS-1737978,  DGE-2039542  &  MRI-1828467  and  FAIN  award  No  1906630;  and an  IBM  faculty award (Research).

The research reported herein was supported in part by NIH award 1R01HG006844, NSF awards CICI-1547324, IIS-1633331, CNS-1837627, OAC-1828467, DMS-1737978, DGE-2039542, MRI-1828467, ARO award W911NF-17-1-0356, FAIN award number 1906630, and IBM faculty award (Research).



%\bibliographystyle{IEEEtran}



% \end{thebibliography}
%\newpage
\bibliographystyle{IEEEtran}
\bibliography{main.bib}

\end{document}